\documentclass[useAMS,usenatbib,twocolumn]{mn2e}

\usepackage{amssymb,amsmath}
\usepackage{graphicx}
\usepackage{url}

\newcommand{\be}{\begin{equation}}
\newcommand{\ee}{\end{equation}}

\title[Polytropic filaments]{Polytropic models of filamentary interstellar clouds -- I \\
Structure and stability} 

\author[Toci \& Galli]{Claudia Toci$^1$\thanks{E-mail: claudia@arcetri.astro.it}, 
Daniele Galli$^2$ \\
$^1$Dipartimento di Fisica e Astronomia, Universit\`a degli Studi di Firenze, Via G. Sansone 1, 
I-50019 Sesto Fiorentino, Italy \\
$^2$INAF-Osservatorio Astrofisico di Arcetri, Largo E. Fermi 5, I-50125 Firenze, Italy}

\begin{document}

\maketitle

\begin{abstract} 

\noindent The properties of filamentary interstellar clouds observed
at sub-millimetre wavelengths, especially by the {\em Herschel Space
Observatory}, are analysed with polytropic models in cylindrical symmetry.
The observed radial density profiles are well reproduced by negative-index
cylindrical polytropes with polytropic exponent $1/3\lesssim
\gamma_{\rm p} \lesssim 2/3$ (polytropic index $-3\lesssim n \lesssim
-3/2$), indicating either external heating or non-thermal pressure
components. However, the former possibility requires unrealistically
high gas temperatures at the filament's surface and is therefore very
unlikely. Non-thermal support, perhaps resulting from a superposition of
small-amplitude Alfv\'en waves (corresponding to $\gamma_{\rm p}=1/2$), is a
more realistic possibility, at least for the most massive filaments. If
the velocity dispersion scales as the square root of the density (or
column density) on the filament's axis, as suggested by observations,
then polytropic models are characterised by a uniform
width. The mass per unit length of pressure-bounded cylindrical polytropes
depends on the conditions at the boundary and is not limited as in the
isothermal case. However, polytropic filaments can remain stable to
radial collapse for values of the axis-to-surface density contrast
as large as the values observed only if they are supported by a 
non-isentropic pressure component.

\end{abstract} 

\begin{keywords}
ISM: clouds -- instabilities
\end{keywords}

\section{Introduction}
\label{intro}

The filamentary structure of molecular clouds has recently received
considerable attention, especially thanks to the high sensitivity and
dynamic range of images obtained at sub-millimetre wavelengths by the {\em
Herschel Space Observatory}. The observed filaments typically represent
enhancements by a factor of $\sim 10^2$ in volume density (or by a factor of $\sim 10$
in column density) with respect to
the ambient medium of the molecular cloud, extending over $\sim$~pc scales
and often forming complex networks
(Andr\'e et al.~2010, Molinari et al.~2010).  From a theoretical
point of view, the origin of interstellar filaments still needs to be
fully understood.  It is debated whether filaments are stagnation regions
formed either at the intersections of planar shocks (Padoan et al.~2001), or
by the collapse and fragmentation of self-gravitating gaseous sheets
(Burkert \& Hartmann~2004) perhaps mediated by magnetic fields (Miyama,
Narita \& Hayashi~1987a,b; Nagai, Inutsuka \& Miyama~1998; Van Loo, Keto
\& Zhang~2014), or they are long-lived features of the flow resulting from
hierarchical fragmentation (G\'omez \& V\'azquez-Semadeni~2013), or they
are formed by turbulent shear and maintained coherent by magnetic stresses
(Hennebelle~2013).  Remarkably, the observed properties of individual
filamentary clouds are well characterised and rather uniform, at least 
for filaments located in nearby molecular clouds (mostly in the Gould's Belt, see 
Andr\'e~2013 for a review).  Their density profiles in the radial
direction, perpendicular to the filament's axis, are characterised
by a flat-density inner part of size $\sim \varpi_{\rm flat}$ and
a power-law envelope extending to an outer radius $\sim 10\,\varpi_{\rm flat}$,
where the filaments merge with the surrounding ambient medium.  The size
of the flat-density region appears to have a uniform value $\varpi_{\rm
flat}=(0.03\pm0.02)$~pc despite a variation of the central column
density $N_c$ of about 3 orders of magnitude, from $\sim 10^{20}$ to
$\sim 10^{23}$~cm$^{-2}$ (Arzoumanian et al.~2011, hereafter A11).

A convenient parametrisation of the radial density profile that reproduces 
these basic features is the softened power-law profile
\be
\rho(\varpi)=\frac{\rho_c}{[1+(\varpi/\varpi_{\rm flat})^2]^{\alpha/2}},
\label{soft}
\ee
where $\rho_c$ is the central density and $\alpha$ is a parameter.
If $\alpha=4$, eq.~(\ref{soft}) is an exact solution of the equation
of hydrostatic equilibrium for a self-gravitating isothermal cylinder,
hereafter referred to as the Stod\'o\l kiewicz--Ostriker density profile (Stod\'o\l
kiewicz~1963, Ostriker~1964a).  In this case $\varpi_{\rm flat}=(2a^2/\pi
G\rho_c)$, where $a$ is the isothermal sound speed. 
The isothermal cylinder has infinite radius, but finite mass per unit length,
\be
\mu_{\rm iso}=\frac{2a^2}{G}=16.5\left(\frac{T}{10~{\rm K}}\right)\, 
\mbox{$M_\odot$~pc$^{-1}$}.
\label{linemass_iso}
\ee
However, the power-law slope $\alpha$
measured in a sample of filaments in the IC5146, Aquila and
Polaris clouds, mapped by the {\em Herschel} satellite,
is significantly different from $\alpha=4$: on average,
$\alpha=1.6\pm 0.3$ (A11). Thus, the possibility
that the gas in these filaments obeys a non-isothermal equation of state, 
and the implications of relaxing the hypothesis of thermal support, should be explored.

A fundamental difference between the behaviour of {\em isothermal} spherical and cylindrical
interstellar clouds with respect to gravitational collapse was pointed out by McCrea~(1957):
while for a spherical cloud of given mass and temperature there is a maximum value
of the external pressure for which an equilibrium state is possible (the Bonnor-Ebert criterion), a cylindrical cloud can be 
in equilibrium for any value of the external pressure, provided its mass per unit length is smaller than a 
maximum value. 
This led McCrea~(1957) to conclude that filamentary (or sheet-like) clouds must first break up 
into fragments of roughly the same size in all directions before gravitational collapse 
(and therefore star formation) can take place. However, as we 
will argue in Sect.~\ref{thermal}, filamentary clouds are unlikely to be thermally supported, and 
their radial density profiles are well reproduced by assuming an equations of state ``softer''
than isothermal. In this case, as shown by Viala \& Horedt~(1974a) the behaviour of cylindrical 
clouds with respect to gravitational instability becomes essentially analogous to that of spherical clouds
(see Sect.~\ref{stability}).

The goal of this paper is to analyse the radial density profiles of
filamentary clouds, their stability with respect to collapse, and to
derive from the observed properties some conclusions on the relative
importance of various mechanisms of radial support (or confinement)
of these clouds. In particular, we analyse the stability of filamentary
clouds following ideas explored in theoretical studies of spherical clouds
by McKee \& Holliman~(1999) and Curry \& McKee~(2000), stressing the need 
for non-isentropic models to account for the observed large density contrasts.
As in the case of spherical polytropes, the stability properties of polytropic cylinders depend on the polytropic
exponent $\gamma_{\rm p}$, that characterises the spatial properties of the
filament, and on the adiabatic exponent $\gamma$, that determines
the temporal response of the cloud to adiabatic perturbations.

The paper is organised as follows:
in Sect.~\ref{radial} we analyse the
radial density profiles of filamentary clouds on the basis of polytropic
cylindrical models; in Sect.~\ref{support} we compare the role of thermal
and non-thermal pressure in supporting the cloud against its self-gravity;
the stability to radial collapse of filaments of increasing mass per unit
length under fixed external pressure is analysed in Sect.~\ref{stability};
finally, in Sect.~\ref{conclusions} we summarise our conclusions.
In this paper we focus on unmagnetised filaments. However, 
we consider particular forms of the equation of state that may simulate
the effects of a large-scale or wavelike magnetic field on the cloud's
structure.  Magnetised models are presented in a companion paper (Toci \&
Galli~2014b, hereafter Paper~II).  

\section{Radial density profiles of filamentary clouds}
\label{radial}

\subsection{Basic equations}
\label{basic}

Neglecting magnetic fields, the structure and
evolution of a self-gravitating filament is governed by the force equation
\be
({\bf u}\cdot\nabla){\bf u}=
-\nabla V-\frac{1}{\rho}\nabla p,
\label{force}
\ee
and Poisson's equation
\be
\nabla^2 V=4\pi G\rho,
\label{pois}
\ee
where ${\bf u}$ is the gas velocity, $V$ is the gravitational potential,
and $p$ is the gas pressure. The left-hand side term in eq.~(\ref{force}) represent the effects of
dynamical motions on the momentum balance.  These include the laminar
and turbulent flows associated to the formation of the filament
and/or produced by the gravitational field of the filament itself.
In a cylindrical coordinate system
with the $z$ axis along the filament's axis and the $\varpi$ axis in
the radial direction, assuming azimuthal symmetry ($\partial/\partial\varphi=0$), and neglecting
rotation, the radial component of the left-hand side of eq.~(\ref{force}) reads
\be
({\bf u}\cdot\nabla){\bf u}=
\left(u_\varpi \frac{\partial u_\varpi}{\partial\varpi}
+u_z\frac{\partial u_\varpi}{\partial z}\right) \, {\hat{\bf e}}_\varpi.
\label{vterm}
\ee
The first term in eq.~(\ref{vterm})
represents a ram-pressure compressing the filament. 
If the filament is building mass by accretion from the surrounding
medium, then $u_\varpi$ is negative and decreases inward ($u_\varpi=0$
by symmetry on the filament's axis).  If $u_\varpi$ becomes
subsonic inside the filament, where ${\bf u}$ is expected to be mostly
parallel to the filament's axis (as e.g. in the simulations of G\'omez \& V\'azquez-Semadeni~2013), 
then the internal pressure dominates over the
accretion ram pressure.  Accretion ram-pressure can be neglected
in the central parts of a filament, although it may play a role in the
envelope\footnote{However, for a Larson-Penson type of accretion, $u_\varpi$
approaches a constant value at large radii (Kawachi \& Hanawa~1992)
and the accretion ram-pressure drops to zero. The second term in
eq.~(\ref{vterm}) is negligible if the accretion velocity $u_\varpi$ does
not change significantly along the filament, and vanishes in cylindrical
symmetry ($\partial/\partial z=0$).}. 
Thus, a description of the structure of filamentary
clouds in terms of hydrostatic equilibrium models does not necessarily
imply that the velocity field is zero everywhere.
Of course the velocity term in eq.~(\ref{force}) cannot be ignored
during the growth of the varicose (or sausage) gravitational instability when
significant radial and longitudinal gas flows can occur (see e.g.Gehman, Adams \&
Watkins~1996). These motions may lead to the formation of dense prestellar cores 
as observed e.g. in the SDC13 infrared dark cloud (Peretto et al.~2014). 

\subsection{Isothermal models}
\label{isothermal}

Observations of limited spatial extent of intensity profiles have
been successfully modelled with isothermal cylinders. For example,
radial density profiles derived from molecular line emission in L1517
(Hacar \& Tafalla~2011), and from the 850~$\mu$m emission in the
filamentary dark cloud G11.11-0.12 (Johnstone et al.~2003) are compatible with the
Stod\'o\l kiewicz-Ostriker density profile up to $\sim 0.2$~pc. Fischera
\& Martin~(2013b) have successfully modelled the surface brightness
profiles of 4 filaments observed by {\em Herschel} in the IC5146 region with
truncated isothermal cylinders, limiting their analysis to $\sim
1^\prime$ radial distance from the emission peak on both sides of
the filament (corresponding to 0.13~pc at the distance of 460~pc).
Over this radial extent, the column density profiles obtained by 
eq.~(\ref{soft}) with $\alpha=2$
or $\alpha=4$, or by a gaussian profile, are all indistinguishable
(see e.g. Fig.~4 of A11).  The large dynamic range
allowed by the {\em Herschel} Space Observatory has made possible to map
the sub-millimetre emission of interstellar filaments up to the radial
distances from the filament's axis where the structures merge with the
ambient medium ($\sim 0.4$~pc for B211/213, Palmeirim et al.~2013; $\sim
1$~pc for IC5112, A11). At large radii, deviation
of the observed density profiles from the St\'odo\l kiewicz-Ostriker
profile become evident, and the observations are generally not well
reproduced by isothermal cylinders. First, as already mentioned in Sect.~\ref{intro}, 
the density profiles at large radii are characterised by power-law exponents $\alpha$ close to
$\sim 2$, rather than $4$; second, the mass per unit length is in some
cases larger than the maximum value allowed for an isothermal cylinder
(eq.~\ref{linemass_iso}).  These aspects will be considered in the
following sections.

\subsection{Polytropic models}
\label{polytropic}

A more general class of hydrostatic models for filamentary clouds
is represented by polytropic cylinders (Ostriker~1964a, Viala \&
Horedt~1974a), in which the gas pressure (arising from thermal or
non-thermal motions) is parametrised by a polytropic equation of state,
\be 
p=K\rho^{\gamma_{\rm p}},
\label{poly}
\ee 
where $\rho$ is the gas density, $K$ a constant and $\gamma_{\rm p}$ is
the polytropic exponent. The constant $K$ is a measure of the cloud's
entropy (for an isothermal gas $K=a^2$, where $a$ is the isothermal sound
speed). The polytropic exponent is usually written as $\gamma_{\rm p}=1+1/n$,
where the polytropic index $n$ can take values in the range $n\le -1$ or
$n>0$ (the range $-1<n<0$ corresponds to negative values of $\gamma_{\rm p}$ and
is therefore unphysical).  For $0\leq \gamma_{\rm p}\leq
1$ ($n\leq -1$) polytropic cylinders have infinite radii and infinite mass per unit length,
whereas for $\gamma_{\rm p}> 1$ ($n>0$) the density and pressure become zero
at some finite radius and therefore have finite masses per unit length.
For $\gamma_{\rm p}=1$ ($n \rightarrow -\infty$) the gas is isothermal,
whereas for $\gamma_{\rm p}\rightarrow 0$ ($n=-1$) the equation of state
becomes ``logatropic'', $p\propto\ln\rho$. This latter form was first
used by Lizano \& Shu (1989) to model the non-thermal support in
molecular clouds associated to the observed supersonic line widths
(see also McLaughlin \& Pudritz~1996, 1997).  Logatropic cylinders have
infinite radius and infinite mass per unit length (Gehman et al.~1996,
Fiege \& Pudritz~2000).  Negative index polytropes ($0\leq \gamma_{\rm p} <
1$) were first proposed as models for thermally-supported interstellar
clouds heated by an external flux of photons or cosmic rays (Viala~1972,
Shu et al.~1972, de Jong et al. 1980). On the other hand, Maloney~(1988) interpreted the
polytropic temperature $T\propto (p/\rho)^{1/2}$ as a measure of
the contribution of non-thermal (turbulent) motions to the support of
the cloud.  In this case negative index polytropes reproduce the observed
increase of non-thermal line width with size observed in molecular clouds
(McKee \& Holliman~1999, Curry \& McKee~2000).

With the equation of state (\ref{poly}), the equation of hydrostatic
equilibrium eq.~(\ref{force}) with the advective term set equal to $0$ 
reduces to the standard cylindrical Lane-Emden equation
\be
\frac{1}{\xi}\frac{d}{d\xi}\left(\xi\frac{d\theta}{d\xi}\right)=\pm\theta^n,
\label{le}
\ee
for the non-dimensional density $\theta$ and radius $\xi$ defined by
\be
\varpi=\varpi_0\xi=
\left[\frac{\mp (1+n)K}{4\pi G\rho_c^{1-1/n}}\right]^{1/2} \xi,
\quad
\rho=\rho_c\theta^n.
\label{nondim}
\ee
In eq.~(\ref{le}) and (\ref{nondim}), as well as in the rest of the paper,
the upper (lower) sign is for $0\le\gamma_{\rm p} < 1$ ($\gamma_{\rm p}>1$), and the
subscripts ``$c$'' and ``$s$'' denote values at the center (axis of the
cylinder) and at the surface of the filament, respectively.  Numerical and analytical
solutions of eq.~(\ref{le}) with boundary conditions $\theta=1$ and
$d\theta/d\xi=0$ at $\xi=0$ have been obtained by Viala \& Horedt~(1974b)
for $0<\gamma_{\rm p}<1$, by St\'odo\l kiewicz~(1963) and Ostriker~(1964a)
for $\gamma_{\rm p}=1$, and by Ostriker~(1964a) for $\gamma_{\rm p}>1$. The mass per
unit length $\mu$ is
\be
\mu=2\pi\int_0^{\varpi_s} \rho\varpi\,d\varpi=
\mp\frac{(1+n)K\rho_c^{1/n}}{2G}\xi_s\theta_s^\prime,
\ee
where eq.~(\ref{le}) has been used to simplify the integral.

In order to compare different models for the radial density profiles,
it is necessary to normalise the radial coordinate $\varpi$ to the same
length scale. To the lowest order in a series expansion for small radii,
the density profile of polytropic filaments is
$\rho(\varpi)\approx \rho_c(1-\varpi^2/\varpi_{\rm core}^2+\ldots)$. The ``core
radius'' $\varpi_{\rm core}$ is
\be
\varpi_{\rm core}=\frac{2\varpi_0}{\sqrt{\mp n}}=
\left(\frac{1+n}{n}\right)^{1/2}
\frac{\sigma_c}{(\pi G\rho_c)^{1/2}},
\label{rcore}
\ee
where $\sigma_c=(p_c/\rho_c)^{1/2}$ is the velocity dispersion on
the filament's axis\footnote{A comparison with an analogous series
expansion of the softened power-law profile (eq.~\ref{soft}) leads to
the identification $\varpi_{\rm core}=(2/\alpha)^{1/2}\varpi_{\rm flat}$.
Since $\alpha\approx 2$, $\varpi_{\rm core}\approx \varpi_{\rm flat}$.}.
For the observed value $\sigma_c\approx 0.26$~km~s$^{-1}$ (Arzoumanian et
al.~2013, hereafter A13), using the fiducial value $n_c\approx 2\times 10^4$~cm$^{-3}$,
and setting $n=-2$ the core radius is
\be
\varpi_{\rm core} \approx  0.047 
\left(\frac{\sigma_c}{0.26~\mbox{km~s$^{-1}$}}\right)
\left(\frac{n_c}{2\times 10^4~\mbox{cm$^{-3}$}}\right)^{-1/2}\, {\rm pc}.
\ee
Fig.~1 compares the density profiles of various cylindrical polytropes
of positive and negative index, as function of radius normalised to
$\varpi_{\rm core}$. The longitudinally averaged density profiles of the filaments in
IC5146, given by eq.~(\ref{soft}) with $\alpha=1.6\pm 0.3$ (A11),
are well reproduced by cylindrical polytropes with $1/3\lesssim
\gamma_{\rm p} \lesssim 2/3$ ($-3\lesssim n \lesssim -3/2$) at least over
the observed radial extent of the filaments (from $\varpi\approx
0.1\,\varpi_{\rm core}$ to $\varpi\approx 10\,\varpi_{\rm core}$). 
Overall, the single value $\gamma_{\rm p}\approx 1/2$ ($n=-2$) provides a 
good fit to the data, at least for this sample of filaments. 
The implications of these results are discussed in Sect.~\ref{nonthermal}.

\begin{figure}
\begin{center}
\includegraphics[width=8cm]{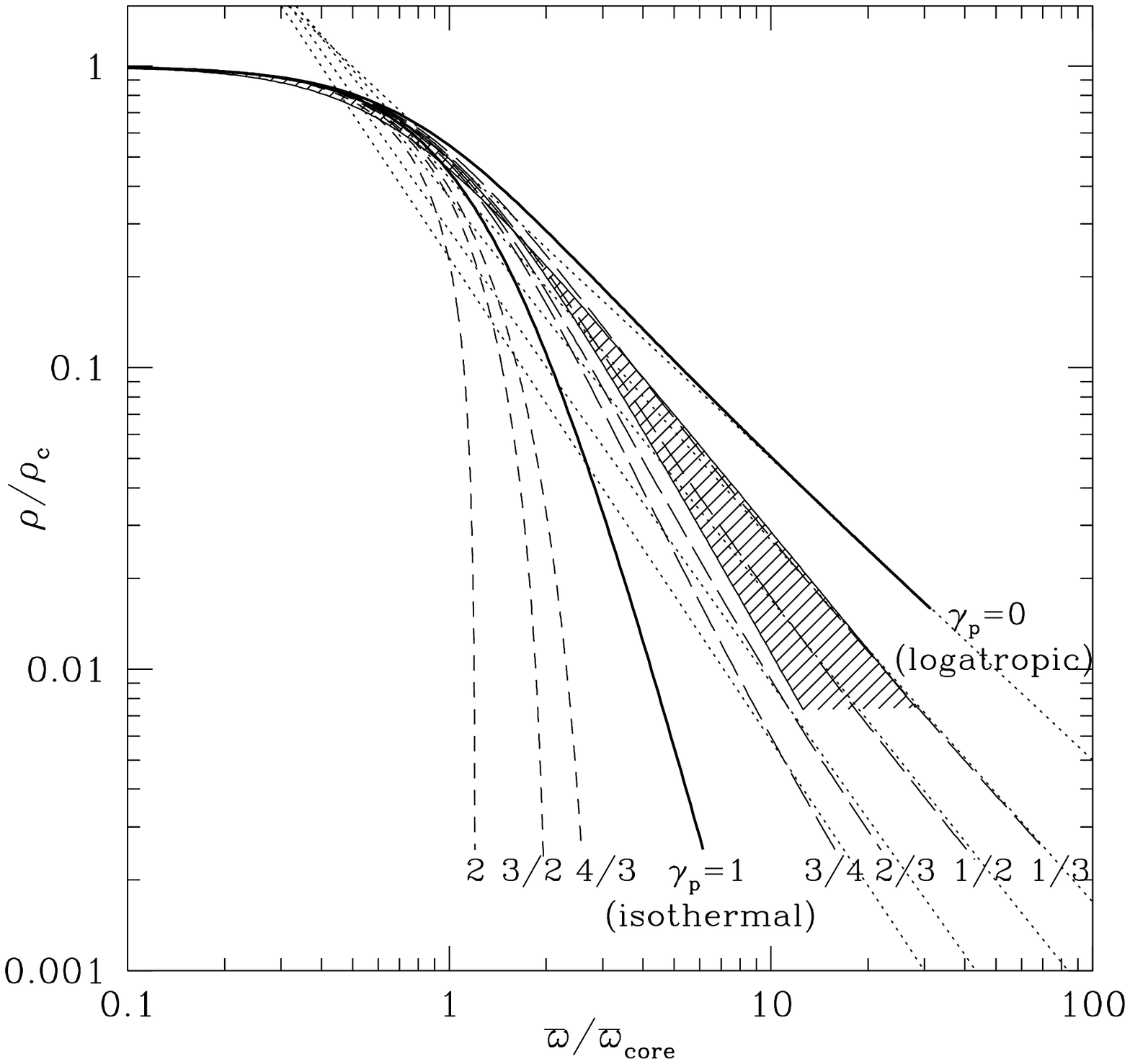}
\caption{Radial density profiles (normalised to the central density
$\rho_c$) of polytropic cylinders with $\gamma_{\rm p}=2$, 3/2, 4/3 ($n=1$, 2 and 3, {\em
short-dashed} lines, from left to right) and $\gamma_{\rm p}=1/3$, 1/2, 2/3 and 3/4 ($n=-3/2$, $-2$, $-3$ and $-4$,
{\em long-dashed} lines, from right to left). The {\em thick solid}
lines show the density profiles of an isothermal ($\gamma_{\rm p}=1$, or
$n=\pm\infty$) and a logatropic ($\gamma_{\rm p}=0$, or $n=-1$) cylinder. {\em
Dotted} lines are the singular solutions given by eq.~(\ref{sing}). The
radius is normalised to the core radius $\varpi_{\rm core}$ defined by
eq.~(\ref{rcore}). The hatched area corresponds to the observed mean
density profile of filaments in IC5146, given by eq.~(\ref{soft}) with
$\alpha=1.6\pm 0.3$.}
\end{center}
\label{fig_profiles}
\end{figure}

\subsection{Power-law behaviour at large radius}
\label{powerlaw}

In addition to regular solutions, eq.~(\ref{le}) also allows singular
(or scale-free) solutions for $0\le\gamma_{\rm p}<1$ ($n<-1$),
characterised by a power-law behaviour
intermediate between $\rho\propto\varpi^{-1}$ (for $\gamma_{\rm p}=0$) and
$\rho\propto\varpi^{-2}$ (for $\gamma_{\rm p}\rightarrow 1$), given by
\be
\rho(\varpi)=\left[\frac{(1-n)^2 \pi G}{-(1+n)K}\right]^{n/(1-n)}\,\varpi^{2n/(1-n)},
\label{sing}
\ee
(Viala \& Horedt~1974a).
The mass per unit length of the scale-free models is
\begin{eqnarray}
\lefteqn{\mu(\varpi)=(1-n)\pi \left[\frac{(1-n)^2 \pi G}{-(1+n)K}\right]^{n/(1-n)}\,\varpi^{2/(1-n)}} \\ \nonumber
& & = (1-n)\pi\varpi^2\rho (\varpi),
\end{eqnarray}
and approaches the constant value $\mu\rightarrow a^2/G$ if $\gamma_{\rm p}\rightarrow 1$.
 
The scale-free solutions are plotted in Fig.~1 along with the regular
solutions. As shown by Fig.~1, scale-free solutions represent the
asymptotic behaviour around which the regular solutions oscillate with
decreasing amplitude for $\varpi\rightarrow\infty$. This asymptotic
behaviour is the same for both polytropic spheres and cylinders.  However,
while a spherical singular solution exists also for $\gamma_{\rm p}=1$ (the
singular isothermal sphere), this does not happen in cylindrical
geometry\footnote{Conversely, for a logatropic equation of state,
a singular solution exist in cylindrical geometry but not in spherical
geometry.}.  In fact, whereas for spheres the amplitude of the oscillatory
component decreases as $r^{-1/2}$ for $\gamma_{\rm p}=1$, for cylinders
it approaches instead a constant value for $\gamma_{\rm p}\rightarrow 1$, and the period of the oscillation
becomes infinite: the isothermal
cylinder converges to the singular solution eq.~(\ref{sing}) only at
infinite radius. Thus, if a quasi-isothermal filamentary clouds goes
trough an evolutionary stage independent of the initial and boundary
conditions, yet still far from the ultimate equilibrium state (an
intermediate asymptotic, Barenblatt~1979), a radial density profile
closer to $\rho\propto\varpi^{-2}$ rather than $\rho\propto\varpi^{-4}$
should be expected. This happens, for example, in the self-similar collapse 
solutions of quasi-isothermal filaments by Kawachi \& Hanawa~(1998).

\section{Support against gravity}
\label{support}

\begin{figure}
\begin{center}
\includegraphics[width=8cm]{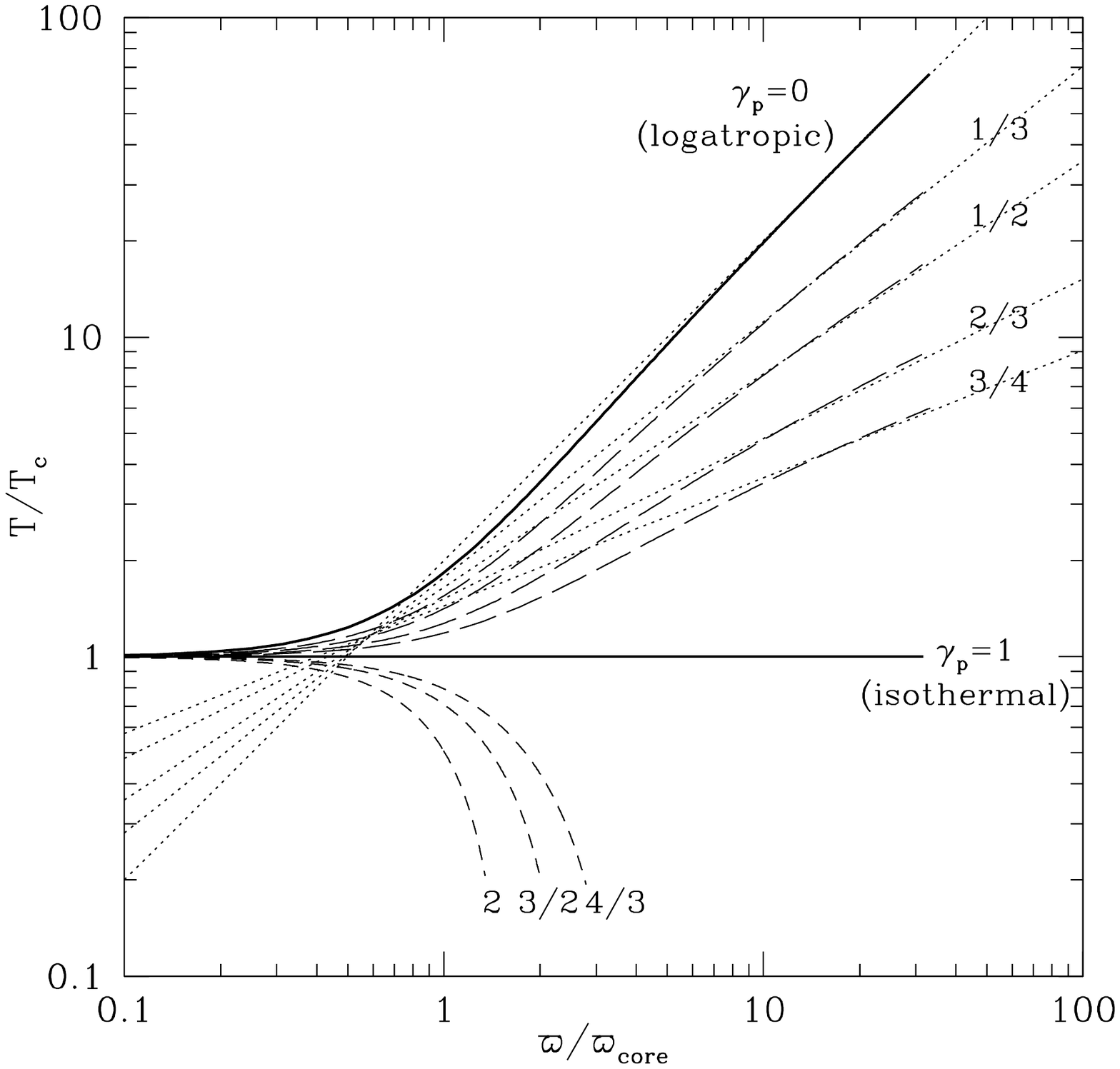}
\caption{Radial profiles of the polytropic temperature $T$ (normalised
to the central temperature value $T_c$) of polytropic cylinders with values 
of $\gamma_{\rm p}$ (or $n$) as in Fig.~1. The
{\em thick solid} lines show the temperature profiles of an isothermal
($\gamma_{\rm p}=1$, or $n=\pm\infty$) and a logatropic ($\gamma_{\rm p}=0$,
or $n=-1$) cylinder.  {\em Dotted} lines correspond to the singular
solutions given by eq.~(\ref{sing}).  The radius is normalised to the
core radius $\varpi_{\rm core}$ as in Fig.~1}
\end{center}
\label{fig_temp} 
\end{figure}

\subsection{Thermal support}
\label{thermal}

The deviations of the observed radial behaviour of the density from an
isothermal St\'odo\l kiewicz-Ostriker profile has been interpreted as
an indication of temperature gradients increasing outwards, resulting
in a larger thermal pressure gradient with respect to an isothermal gas (Recchi,
Hacar \& Palestini~2013). This possibility is supported by the presence of
significant radial gradients in the dust temperature profiles derived from
radiative transfer modelling of the infrared emission (see e.g. Stepnik
et al.~2003). In particular, the dust temperature $T_{\rm d}$ increases
outward from $\sim 10$--12~K on the axis to $\sim 14$~K at $\varpi\approx
0.5$~pc in the B211 filament (Palmeirim et al.~2013), and to $\sim 18$~K
in the L1506 filament (Ysard et al.~2013). Similar (or larger) gradients
are expected in the gas temperature $T_{\rm g}$ as well: in fact,
while some mild coupling of the dust and gas temperatures is possible
at the typical densities on the filament's axis ($\sim 10^4$~cm$^{-3}$),
in general the gas is expected to be significantly hotter than the dust
in the outer regions (see e.g. Galli, Walmsley \& Gon\c calves~2002; Keto \& Caselli~2008).

The polytropic models presented in Sect.~\ref{polytropic} make possible to
quantify the magnitude of the gradient in the gas temperature needed
to reproduce the observed density profiles.  Fig.~2 shows the radial
behaviour of polytropic temperature $T\propto (p/\rho)^{1/2}$ for the
same models shown in Fig.~1. For the range of polytropic exponents
that reproduce the density profiles of filaments in IC5146, the polytropic
temperature increases by a factor $\sim 5$--12 from the filament's axis
to the boundary, assumed to be located at $\rho_s\approx 10^{-2}\rho_c$
or $\varpi_s\approx 10\,\varpi_{\rm core}$ (corresponding to a radius
of about 1~pc, where the filaments merge with the ambient medium). If
the polytropic temperature is identified with the gas temperature, this
implies a temperature at the filament's surface $T_s\sim 70$--170~K,
assuming a central temperature $T_c=14$~K (A11).
Such high temperatures are very unlikely. The observed gradients of gas
temperature in prestellar cores are much shallower (Crapsi et al.~2007),
in agreement with the predictions of theoretical models. Therefore, 
alternatives to thermal pressure must be sought.

\subsection{Non-thermal support}
\label{nonthermal}
 
Turbulence and magnetic fields, either large-scale or wavelike, can
contribute to the pressure supporting the filament.  If approximated
as isotropic pressure components, their effects can be modelled
with appropriate polytropic laws.  For example, in the limit of small
amplitude, small wavelength, and negligible damping, Alfv\'en waves behave
as a polytropic gas with $\gamma_{\rm p}=1/2$ (Wal\'en~1944), a value consistent
with the observations, as shown in Sect.~\ref{polytropic}. Thus, the
filamentary clouds observed by {\em Herschel} may be supported radially
by non-thermal motions associated to Alfv\'enic ``turbulence'', i. e. 
a superposition of hydromagnetic waves (Fatuzzo \& Adams~1993, McKee \&
Zweibel~1995).  

If small-amplitude Alfv\'en waves (modelled with a $\gamma_{\rm p}=1/2$ polytropic
law) dominate the pressure, observed molecular transitions should
be characterised by a non-thermal line width increasing roughly by
a factor $\sim 3$ from the axis to the filament boundary, following
approximately a $\rho^{-1/4}$ (or $\varpi^{1/3}$) dependence at large
radial distances.  However the available data do not allow any firm
conclusion to be drawn on the magnitude and spatial distribution of
non-thermal motions inside filamentary clouds.  Hacar \& Tafalla~(2011)
find that in L1517 the non-thermal line width of molecular transitions
like C$^{18}$O and SO is everywhere subsonic ($\sigma_{\rm nt}<a$)
and very uniform, typically $\sigma_{\rm nt}=0.1\pm 0.04$~km~s$^{-1}$
across the sampled region.  Li \& Goldsmith (2012) find that the velocity
dispersion on the axis of the B213 filament is slightly supersonic
($\sigma_{\rm nt}\approx 0.3$~km~s$^{-1}$).  Millimetre line studies
indicate that self-gravitating filaments have intrinsic, suprathermal
linewidths $\sigma_{\rm nt} \gtrsim a$ (A13).  In the
massive filament DR21 ($N_c\approx 10^{23}$~cm$^{-2}$, $\mu\approx
4\times 10^3$~$M_\odot$~pc$^{-1}$) Schneider et al.~(2010) find that
the velocity dispersion increases towards the filament's axis, where
it reaches $\sigma_{\rm nt}\approx 1$~km~s$^{-1}$ (see their Fig.~18),
whereas condensations in the filaments are characterised by lower velocity
dispersions. Further observations should explore the spatial distribution
of non-thermal motions in filamentary clouds and the correlation (if any)
of $\sigma_c$ with $\rho_c$.

As shown in Sect.~\ref{radial}, negative-index cylindrical polytropes 
with appropriate values of $\gamma_{\rm p}$ reproduce
the observed radial density profiles of filaments and predict a core radius 
$\varpi_{\rm core}\propto \sigma_c/\rho_c^{1/2}$. This result is consistent with the
observed uniformity of filament widths if $\sigma_c$ scales as the
square root of the central density, $\sigma_c\propto \rho_c^{1/2}$.
Since the central column density is $N_c\propto\rho_c\varpi_{\rm core}$,
it follows that $\sigma_c \propto N_c^{1/2}$, since $\varpi_{\rm
core}$ is constant.  Observationally, filaments with central column
densities above $\sim 10^{22}$~cm$^{-2}$ follow this trend (A13).
Theoretically, the relation $\sigma\propto \rho^{1/2}$
seems to characterise the behaviour of the turbulent pressure during
the relaxation processes leading to virialization in a strongly
self-gravitating collapse flow, according to the numerical simulations
of V\'azquez-Semadeni, Cant\'o \& Lizano~(1998). This could be an indication that,
at least in the more massive filaments, the gas in the central parts is
still undergoing turbulent dissipation (perhaps following accretion, 
Hennebelle \& Andr\'e~2013).
Numerical simulations and analytic considerations show that the polytropic exponent 
of magnetohydrodynamic turbulence depends on the dominant wave mode 
via the Alfv\'en Mach number $M_{\rm A}$, ranging from $\gamma_{\rm p}\approx 1/2$ at 
low $M_{\rm A}$, where the slow mode dominates, to $\gamma_{\rm p}\approx 2$ at 
large $M_{\rm A}$, where the slow and fast mode are comparable
(Passot \& V\'azquez-Semadeni~2003).
Thus,  a picture of magnetohydrodynamic turbulence in terms of small-amplitude Alfv\'en 
waves is clearly an oversimplification.


\section{Radial stability of polytropic filaments}
\label{stability}

Cylindrical polytropes are known to be unstable to longitudinal
perturbations of wavelength larger than some critical value. This
varicose (or sausage) gravitational instability (Ostriker~1964b;
Larson~1985; Inutsuka \& Miyama~1992; Freundlich, Jog \& Combes~2014)
and its magnetic variant (Nagasawa~1987; Nakamura et al.~1993; Gehman
et al.~1996; Fiege \& Pudritz~2000b) produces the fragmentation of a
filamentary cloud in a chain of equally spaced dense cores, as observed
in some cases, and represents therefore a promising mechanism for star
formation. However, it is important to assess first the conditions for
radial stability. i.e. with respect to collapse to a line mass, in analogy
with the Bonnor-Ebert stability criterion for spherical polytropes.
For observed filaments, stability considerations are usually based on
a comparison with the mass per unit length of the isothermal cylinder,
$\mu_{\rm iso}$ (eq.~\ref{linemass_iso}). However, as mentioned in the 
Introduction, the stability properties of an isothermal cylinder
are different from those of polytropic cylinders with $\gamma_{\rm p}<1$,
essentially because its mass per unit length approaches the finite value
$\mu^{\rm iso}$ as the radius of the cylinder increases to infinity,
whereas for $\gamma_{\rm p}<1$ the mass increases with radius.  As a consequence,
an isothermal filament is always radially stable: if the pressure
$p_s$ exerted over an isothermal cylinder with fixed
$\mu<\mu_{\rm iso}$ is gradually increased, the filament contracts,
reducing its radius $\varpi_s$ and core radius $\varpi_{\rm core}$ as
$p_s^{-1/2}$, and increasing its central density $\rho_c$ as $p_s^{-1}$,
but otherwise maintaining the same shape of the density profile.  Conversely, for
$0\le\gamma_{\rm p}<1$, also cylindrical polytropes become unstable if the external
pressure becomes larger than some critical value. The instability extends
to $\gamma_{\rm p}=4/3$ for spheres (the classical Bonnor-Ebert instability)
but not for cylinders.

The stability of polytropic cylindrical clouds to radial perturbations
can be determined by solving the equation of radial motion for small 
perturbations about equilibrium, first derived for spherical
clouds by Eddington~(1926). For cylindrical clouds it becomes (Breysse, 
Kamionkowski \& Benson~2014)
\be
\frac{d^2 h}{d\varpi^2}+\frac{3-4q}{\varpi}
\frac{dh}{d\varpi}+\left[\frac{\omega^2}{f^2}
+8\left(\frac{1}{\gamma}-1\right)q\right]\frac{h}{\varpi^2}=0,
\label{osc}
\ee
where $h=\delta\varpi/\varpi$ is the relative amplitude of the
perturbation, $\omega$ is the frequency of the oscillations, $\gamma$
is the adiabatic exponent, and we have defined
\be
q\equiv\frac{G\mu\rho}{2p}=-\frac{(1+n)\xi\theta^\prime}{4\theta},
\ee 
and 
\be 
f\equiv\frac{1}{\varpi}\left(\frac{\gamma p}{\rho}\right)^{1/2}
=\frac{(4\pi G\rho_c)^{1/2}}{\xi}\left(\frac{\mp\gamma\theta}{1+n}\right)^{1/2}.
\ee
In deriving eq.~(\ref{osc}), the simplifying assumption has been made that the
perturbations occur adiabatically, $\delta p/p=\gamma\,\delta\rho/\rho$. It is important to notice
that the adiabatic exponent $\gamma$ determining the response of the cloud to small
perturbations is not necessarily equal to the polytropic exponent characterising the 
equilibrium structure discussed in Sec.~\ref{radial}. Only if the perturbation occurs on a time 
much longer than the characteristic time for internal redistribution of entropy, the adiabatic 
exponent $\gamma$ is equal to $\gamma_{\rm p}$ (see examples and discussion in 
Sect.~\ref{nonisentropic}). 

\subsection{Isentropic filaments}
\label{isentropic}

\begin{figure}
\begin{center}
\includegraphics[width=8cm]{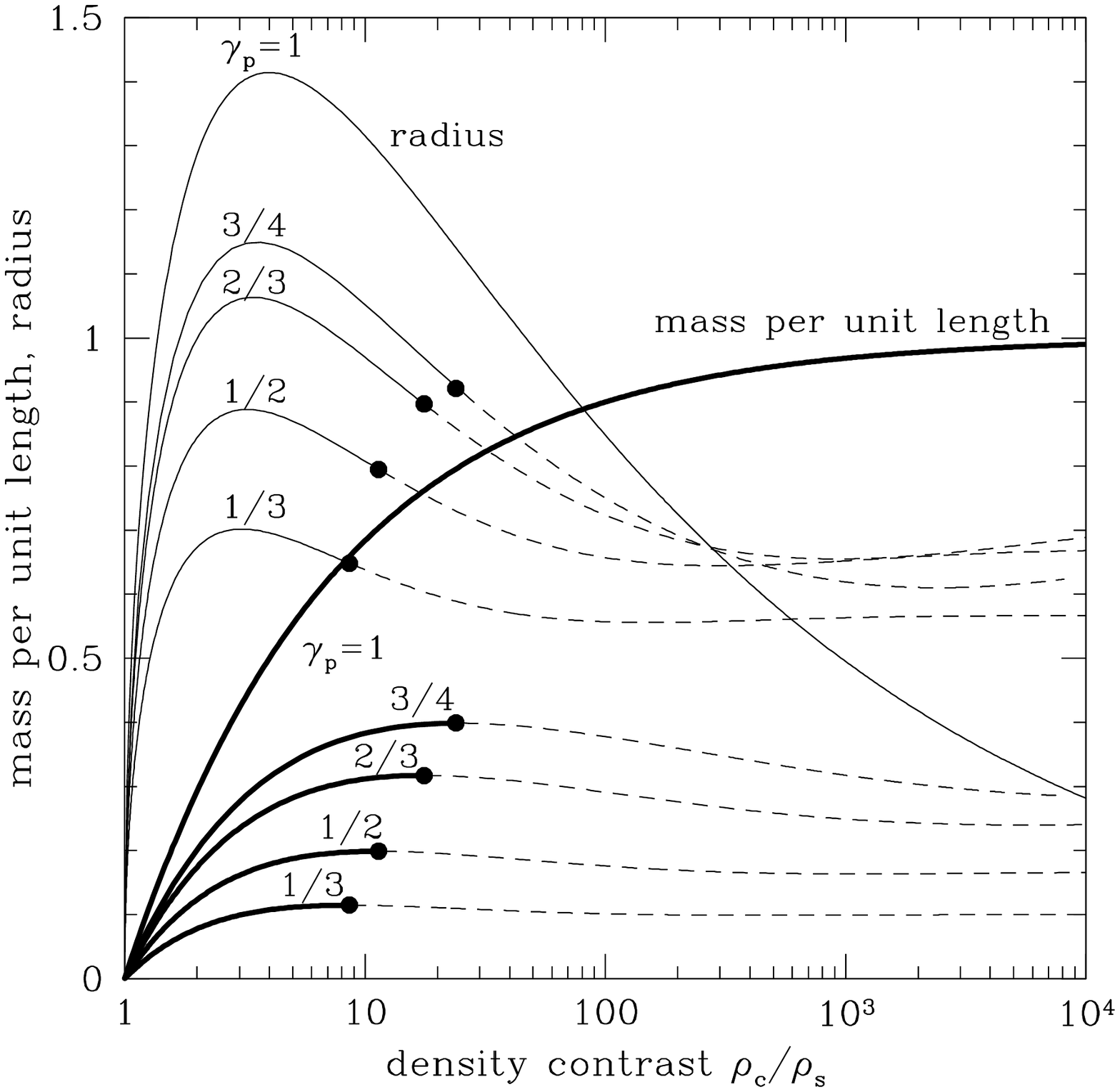}
\caption{Mass per unit length
$\mu$ ({\em thick} curves) and radius $\varpi_s$ ({\em thin} curves) 
of cylindrical polytropes bounded a fixed
external pressure as function of the density contrast $\rho_c/\rho_s$.
The cases shown are, from bottom to top, $\gamma_{\rm p}=1/3$, $1/2$, $2/3$,
$3/4$ and $1$ ($n=-1.5$, $-2$, $-3$, $-4$ and $\-\infty$). Dots on each
curve indicate critical points. The stable and unstable parts of each
sequence are shown by {\em solid} and {\em dashed} curves, respectively.
The radius $\varpi_s$ is in units of $[p_s/(4\pi G\rho_s^2)]^{1/2}$,
the mass per unit length $\mu$ in units of $2p_s/G\rho_s$.}
\end{center}
\label{fig_critical} 
\end{figure}

\begin{table}
\caption{Critical points for isentropic cylindrical polytropes.}
\label{tab_f}
\begin{tabular}{lllll}
\hline
$n$ & $\gamma_{\rm p}$ & $\xi_{\rm cr}$ & $(\rho_c/\rho_s)_{\rm cr}$ & $q_{\rm cr}$ \\
\hline
$-1$     & $0$         & $6.62$   & $6.05$ & $0$      \\
$-1.01$  & $0.0099$    & $6.59$   & $6.10$ & $0.0272$ \\
$-1.5$   & $1/3$       & $5.52$   & $8.61$ & $0.115$   \\
$-2$     & $1/2$       & $4.93$   & $11.4$ & $0.199$   \\
$-3$     & $2/3$       & $4.28$   & $17.6$ & $0.317$   \\
$-4$     & $3/4$       & $3.92$   & $23.9$ & $0.399$   \\
$-5$     & $0.8$       & $3.68$   & $32.8$ & $0.459$   \\
$-10$    & $0.9$       & $3.13$   & $80.0$ & $0.626$  \\
$-20$    & $0.95$      & $2.76$   & $228$  & $0.752$   \\
$-30$    & $0.967$     & $2.60$   & $441$  & $0.812$   \\
$-40$    & $0.975$     & $2.50$   & $701$  & $0.846$  \\
$-\infty$ & $1$        &          & $\infty$ & $1$     \\        
\hline
\end{tabular}
\end{table}

We first consider isentropic clouds, in which the entropy is both
spatially uniform and constant during an adiabatic perturbation, and set
$\gamma=\gamma_{\rm p}$.  To determine the condition of marginal stability, we
set $\omega=0$ and we solve eq.~(\ref{osc}) with the boundary condition
$dh/d\xi=0$ at $\xi=0$ in order for $h$ to remain finite on the axis
(since eq.~\ref{osc} is linear and homogeneous, the value of $h$ at
$\xi=0$ is arbitrary). For any fixed value of the polytropic exponent
$\gamma_{\rm p}$, the critical point $\xi_{\rm cr}$ can be found determining
the radius at which the Lagrangian variation in the pressure at the
boundary vanishes,

\be
\left(\frac{\delta p}{p}\right)_{\xi=\xi_{\rm cr}}=
-\gamma\left(2h+\varpi\frac{dh}{d\varpi}\right)_{\xi=\xi_{\rm cr}}=0.
\ee
If $\xi>\xi_{\rm cr}$, the filament is unstable to radial collapse.
At the critical point the density contrast  is $(\rho_c/\rho_s)_{\rm
cr}=\theta^{-n}_{\rm cr}$
and the mass per unit length is 
\be
\mu_{\rm cr}=q_{\rm cr}\left(\frac{2p_s}{G\rho_s}\right),
\label{mu_cr}
\ee
where
\be
q_{\rm cr}=-\frac{(1+n)\xi_{\rm cr}\theta^\prime_{\rm cr}}{4\theta_{\rm cr}}.
\ee
The values of $\xi_{\rm cr}$, $(\rho_c/\rho_s)_{\rm cr}$, and $q_{\rm cr}$
for different polytropes are listed in Table~\ref{tab_f}. For the same
value of the ratio $p_s/\rho_s$, the marginally stable configuration
with the largest mass per unit length is the isothermal filament with
$\gamma_{\rm p}=1$, for which $q_{\rm cr}=1$ and $\mu_{\rm cr}=\mu_{\rm
iso}$. At the opposite end, the logatropic filament with $\gamma_{\rm p}=0$
has $q_{\rm cr}=0$. Thus, for fixed values of the surface pressure and
density, filaments with increasingly ``softer'' equations of state can
support less and less mass per unit length, as in the case of spherical
polytropes (McKee \& Holliman~1999).

Fig.~3 shows the radius and the mass per unit length of
cylindrical polytropes with various values of $\gamma_{\rm p}$
between $\gamma_{\rm p}=1/3$ and 1 (from $n=-3/2$ to $-\infty$) as function
of $\rho_c/\rho_s$ and the position of the critical point on both sets
of curves.  The stability properties of polytropic filaments with $0\le
\gamma_{\rm p}< 1$ for the same value of the entropy parameter $K$ and the ratio
$p_s/\rho_s$ are qualitatively similar: increasing $\rho_c/\rho_s$ the
filament first expands then contracts, until the filament becomes unstable
when $\rho_c/\rho_s$ becomes larger than the critical value listed in
Table~\ref{tab_f}.  Equilibria also exist above this critical value,
but they are unstable to radial collapse. The instability occurs for
increasingly larger values of $\rho_c/\rho_s$ when $\gamma_{\rm p}$ increases
(for $\gamma_{\rm p}=1$, the critical point is at $\xi_{\rm cr}=\infty$). 

The problem here is that several filamentary clouds observed by {\em
Herschel} have mass per unit length in excess of $\mu_{\rm iso}$,
if the latter is computed with $a^2$ corresponding to the measured
central temperature $\sim 10$~K.  Although the fact that prestellar
cores are preferentially found in filaments with $\mu>\mu_{\rm iso}$
is considered a signature of gravitational instability (see, e.g.,
Andr\'e et al.~2010), it is difficult to justify the formation by
accretion (or by other processes) of isothermal filaments with mass
per unit line larger than $\mu_{\rm iso}$.  Consider for example the
evolution of an isothermal filament with $\mu<\mu_{\rm iso}$, bounded by an
external constant pressure $p_s=a^2 \rho_s$, slowly increasing its
mass per unit length while keeping its temperature uniform and constant
with time.  As $\mu$ increases, the filament becomes more and more
centrally condensed, its density contrast $\rho_c/\rho_s$ increasing as
$(1-\mu/\mu_{\rm iso})^{-2}$.  At the same time, the flat core region
shrinks as $(1-\mu/\mu_{\rm iso})$, and the outer radius first expands
then contracts as $[(1-\mu/\mu_{\rm iso})(\mu/\mu_{\rm iso})]^{1/2}$
(Fischera \& Martin~2013a).  As $\mu\rightarrow \mu_{\rm iso}$, the
filament approaches a delta-like line mass of zero radius and infinite
density on the axis. During this evolution the filament is subject
to the varicose instability and can fragment into a chain of cores,
but can never reach a stage with $\mu>\mu_{\rm cr}$.  This is not the case 
for non-isothermal filaments. In fact, it is
reasonable to expect that in actual filaments the ratio $p_s/\rho_s$ in
eq.~(\ref{mu_cr}) is much larger than $a^2$, the value for isothermal gas.
If filamentary clouds are pressure confined, $p_s$ must be equal to the
pressure exerted on the filament by the surrounding intercloud medium,
where turbulent motions are likely to dominate the pressure. 

\subsection{Non-isentropic filaments}
\label{nonisentropic}

If filamentary clouds are well described by
cylindrical polytropes with $1/3\lesssim \gamma_{\rm p}\lesssim 2/3$ as shown
in Sect.~\ref{radial}, their density contrast cannot be larger than
$\rho_c/\rho_s=8.61$--$17.6$ (see Table~\ref{tab_f}) or they would
collapse to a line mass.  However, the observations summarised in
Sect.~\ref{radial} indicate that the density contrasts measured by {\em
Herschel} are of the order of $\sim 100$.  As in the case of spherical
clouds, this limitation is alleviated if the cloud is non-isentropic
($\gamma\neq\gamma_{\rm p}$).

While the assumption of isentropy has been made in most studies
of polytropes, McKee \& Holliman~(1999) and Curry \& McKee~(2000)
showed that it is not generally valid for molecular clouds. In fact, a
significant contribution to the pressure supporting the cloud against its
self-gravity may be provided by non-thermal components whose behaviour
is not isentropic: for example, small-amplitude Alfv\'en waves have
$\gamma_{\rm p}=1/2$ and $\gamma=3/2$ (McKee \& Zweibel~1995). 
In general,
non-isentropic polytropes remain stable for larger density contrasts
than isentropic clouds.  
In practice, the analysis must be limited to values of $\gamma>\gamma_{\rm p}$ since
polytropes with $\gamma<\gamma_{\rm p}$ are convectively unstable according to
the Schwarzschild criterion.  

To obtain the critical point $\xi_{\rm cr}$ of non-isentropic
cylindrical polytropes, eq.~(\ref{osc}) is solved for a fixed
$\gamma_{\rm p}$ and arbitrary $\gamma >\gamma_{\rm p}$.  The results
are shown in Table~\ref{tab_nonis}, listing the values of $\xi_{\rm
cr}$, $(\rho_c/\rho_s)_{\rm cr}$ and $q_{\rm cr}$ for polytropes with
$\gamma_{\rm p}=1/3$, $1/2$ and $2/3$ for various values of 
$\gamma$.  As for the case of spherical polytropes, the critical points
moves to larger and larger values of the density contrast $\rho_c/\rho_s$
as $\gamma$ increases.  At a threshold value $\gamma_\infty$, the critical
point reaches $\xi_{\rm cr}=\infty$ and the density profile approaches
that of a singular polytropic cylinder. The value of $q_{\rm cr}=q_\infty$
at this point can be easily determined substituting eq.~(\ref{sing}) into eq.~(\ref{osc}),
\be
q_\infty=\frac{\gamma_{\rm p}}{2(2-\gamma_{\rm p})}.
\ee
The threshold value of  the adiabatic exponent, $\gamma_\infty$ can also be found
analytically. For a singular polytropic cylinder, eq.~(\ref{osc}) with
$\omega=0$ has constant coefficients, and the characteristic equation
has two real and negative roots if $\gamma$ is larger than
\be
\gamma_\infty= \gamma_{\rm p}(2-\gamma_{\rm p}),
\label{ginf}
\ee
corresponding to $h$ exponentially decreasing with $\xi$. The values
of $q_\infty$ and $\gamma_\infty$ for $\gamma_{\rm p}=1/3$, $1/2$ and $2/3$
are also listed in Table~\ref{tab_nonis}. Non-isentropic polytropes
are more stable than their isentropic counterparts as they can support
larger centre-to-surface density contrasts. For $\gamma>\gamma_\infty$,
polytropic filaments are unconditionally stable for any $\rho_c/\rho_s$.

Fig.~4 summarises the stability properties of
cylindrical polytropes in the $\gamma_{\rm p}$--$\gamma$ plane.
In cylindrical geometry the polytropic exponent $\gamma_{\rm p}=1$ is a critical value that
plays the same role of $\gamma_{\rm p}=4/3$ for spherical polytropes: while 
spheres with $\gamma_{\rm p}>4/3$ are unconditionally stable to small perturbations, 
cylinders become stable already for $\gamma_{\rm p}>1$ (McCrea~1957, Larson~2005). 
The analysis presented in this Section extends the study of the gravitational instability to non-isentropic 
clouds determining the threshold value 
$\gamma_\infty$ for the stability of polytropic cylinders 
as function of the polytropic exponent $\gamma_{\rm p}$.
For example, for $\gamma_{\rm p}=1/2$, a value consistent with the observed 
radial density profiles of filamentary clouds, the threshold value for stability 
(from eq.~\ref{ginf}) is $\gamma_\infty=3/4$. Notice that for a ``soft'' equation of state 
the stability condition is about the same for cylinders and spheres:
for $\gamma_{\rm p}\ll 1$,  a first-order approximation gives $\gamma_\infty
\approx 2\gamma_{\rm p}$ for cylinders and $\gamma_\infty\approx (16/9) \gamma_{\rm p}$ 
for spheres. For larger values of $\gamma_{\rm p}$, cylinders are intrinsically more stable 
than spheres in the $\gamma_{\rm p}$--$\gamma$ plane.  
A pressure-bounded isothermal cylinder, for example, is always stable with respect to an arbitrary
increase in the external pressure, whereas an isothermal sphere is not. For the range 
of polytropic exponents allowed by the observations of the radial density profiles 
($1/3\lesssim \gamma_p \lesssim 2/3$, see Sect.~2), the stability properties of cylindrical 
and spherical clouds are very similar.

\begin{table}
\caption{Stability of non-isentropic cylindrical polytropes.}
\label{tab_nonis}
\begin{tabular}{llll}
\hline
\multicolumn{4}{c}{$\gamma_{\rm p}=1/3$} \\
$\gamma$ & $\xi_{\rm cr}$ & $(\rho_c/\rho_s)_{\rm cr}$  & $q_{\rm cr}$ \\
\hline
$1/3$  &  $5.52$  & $8.61$ & $0.115$ \\
$0.4$  &  $10.3$  & $20.2$ & $0.111$  \\
$\gamma_\infty=5/9$  & $\infty$ & $\infty$ & $q_\infty=1/10$ \\
\hline
\multicolumn{4}{c}{$\gamma_{\rm p}=1/2$} \\
$\gamma$ & $\xi_{\rm cr}$ & $(\rho_c/\rho_s)_{\rm cr}$  & $q_{\rm cr}$ \\
\hline
$1/2$ & $4.93$  & $11.4$  & $0.199$  \\
$0.6$ & $10.7$  & $38.0$  & $0.188$ \\
$0.7$ & $124$   & $1096$ & $0.163$ \\
$\gamma_\infty=3/4$ & $\infty$ & $\infty$  & $q_\infty=1/6$     \\
\hline
\multicolumn{4}{c}{$\gamma_{\rm p}=2/3$} \\
$\gamma$ & $\xi_{\rm cr}$ & $(\rho_c/\rho_s)_{\rm cr}$ & $q_{\rm cr}$  \\
\hline
$2/3$ & $4.28$ & $17.6$ & $0.317$  \\
$0.7$ & $5.18$ & $25.2$ & $0.316$ \\
$0.8$ & $14.1$ & $153$  & $0.282$  \\
$\gamma_\infty=8/9$ & $\infty$ & $\infty$ & $q_\infty=1/4$     \\
\hline
\end{tabular}
\end{table}

\begin{figure}
\begin{center}
\includegraphics[width=8cm]{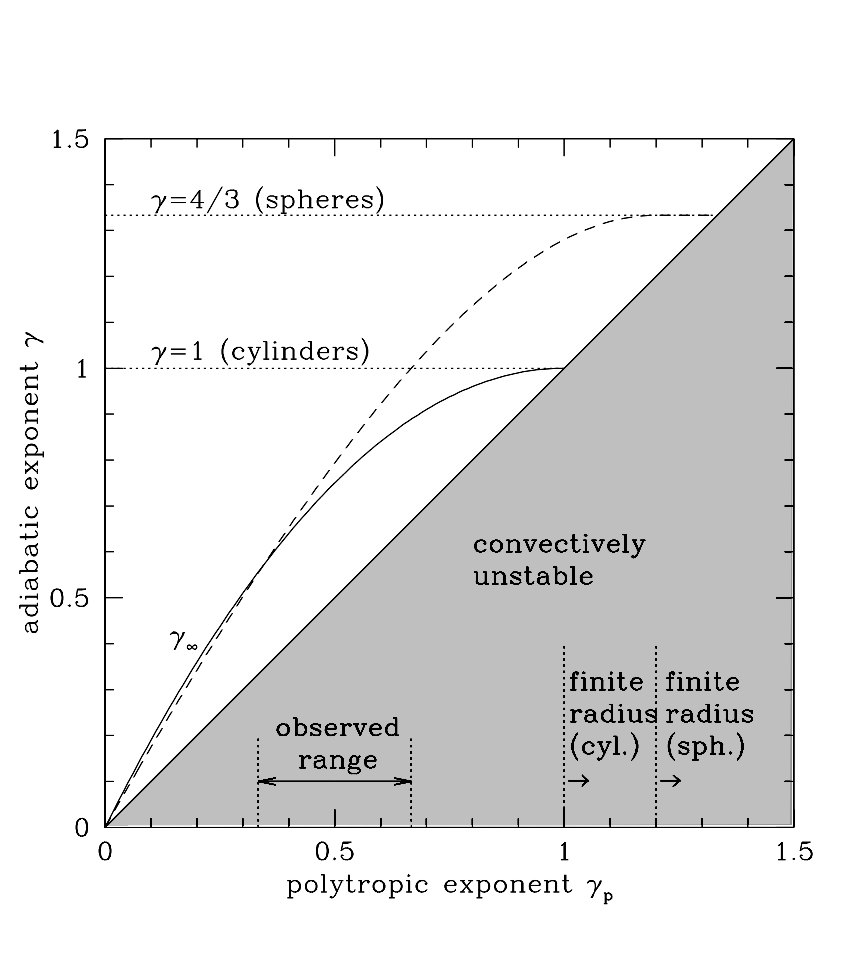}
\caption{Stability properties of cylindrical and spherical polytropes
in the $\gamma_{\rm p}$--$\gamma$ plane. Polytropes in the $\gamma<\gamma_{\rm p}$
region ({\em shaded}) are convectively unstable. On the line $\gamma=\gamma_{\rm p}$, 
polytropic cylinders (spheres) are
isentropic, and become unstable at some finite $(\rho_c/\rho_s)_{\rm
cr}$ if $\gamma<1$ ($\gamma<4/3)$. Above the curve
labelled $\gamma_\infty$ ({\em dashed} for spheres) cylindrical polytropes
are unconditionally stable even for $\rho_c/\rho_s=\infty$.
Cylindrical (spherical) polytropes have finite radii for $\gamma_{\rm p}> 1$
($\gamma_{\rm p}> 6/5$) as indicated by the vertical {\em dotted} lines. The stability
properties of spherical polytropes are from McKee \& Holliman~(1999).}
\end{center}
\label{fig_stability} 
\end{figure}

\section{Conclusions}
\label{conclusions}

The typical core-envelope structure and the uniformity of the
observed properties of filamentary molecular clouds suggests that
their main physical characteristics can be analysed with 
polytropic models in cylindrical symmetry. Isothermal models fail
to reproduce the observed power-law behaviour of the density at radii larger 
than the core radius, and cannot explain the existence of filaments 
with mass per unit length larger than the limiting value for an
isothermal cylinder. Conversely,
the observed radial density profiles of filamentary clouds are well
reproduced by negative-index cylindrical polytropes with $1/3\lesssim
\gamma_{\rm p} \lesssim 2/3$ ($-3\lesssim n \lesssim -3/2$) indicating
either  outward-increasing temperature gradients, or the presence of a
dominant non-thermal contribution to the  pressure.  In the former case,
however, the predicted gas temperature at the filament's surface would
be unrealistically high. Non-thermal support, perhaps in the form of a
superposition of small-amplitude Alfv\'en waves (for which $\gamma_{\rm p}=1/2$)
is an attractive possibility. In addition, the mass per unit length 
of negative-index polytropes is not limited, but depends on the 
pressure and density at the surface, if the filaments are pressure confined
by the ambient medium.

Negative-index cylindrical polytropes have uniform width,
as observed, if the central velocity dispersion $\sigma_c$ is
proportional to the square root of the central density $\rho_c$ (or
the central column density $N_c$), a relation that seems to be satisfied
at least by the most dense filaments (A13) and has
been found in numerical simulations of self-gravitating collapse flows
(V\'azquez-Semadeni et al.~1998).

Outside the core radius, the density profile of polytropic filaments
has often a power-law behaviour and carries important information
on the cloud's thermodynamics and equation of state. Irrespective of
geometry, both spherical and cylindrical polytropes converge at large
radii to the same power-law behaviour in radius with a slope equal
to $-2/(2-\gamma_{\rm p})$, that approaches $-2$ for a quasi isothermal
gas. However, for cylinders, this power-law behaviour is approached
at increasingly larger radii for $\gamma_{\rm p}\rightarrow 1$ (at infinite
radius for  $\gamma_{\rm p}=1$).

Pressure-bounded polytropic cylinders with
$1/3\lesssim\gamma_{\rm p}\lesssim 2/3$ can support a mass per unit length as
large as observed depending on the conditions at the surface.  However,
their density contrast cannot be larger than about a factor of $10$--$20$
if they are isentropic. Like their spherical counterparts (McKee \& Holliman~1999),
non-isentropic cylinders remain stable
at larger density contrasts (in principle even infinite) with respect
to adiabatic pressure perturbations. Since magnetic fields and turbulence 
(modelled here in the very simplified framework of Alfv\'en waves) behave as non-isentropic 
pressure components, isentropic (and, in particular, isothermal) models are inadequate to 
study the structure and the stability properties of filamentary clouds.

\section*{Acknowledgments}

It is a pleasure to acknowledge stimulating discussions with Philippe 
Andr\'e and Mario Tafalla. We also thank an anonymous referee for 
very useful comments that improved the presentation of the paper.

\end{document}